# DNAHLM - DNA sequence and Human Language mixed large language Model

Wang Liang
Huazhong University of Science and Technology, 430070, P.R. China
*To whom correspondence should be addressed. E-mail:wangliang.f@gmail.com

**[Abstract]** There are already many DNA large language models, but most of them still follow traditional uses, such as extracting sequence features for classification tasks. More innovative applications of large language models, such as prompt engineering, RAG, and zero-shot or few-shot prediction, remain challenging for DNA-based models. The key issue lies in the fact that DNA models and human natural language models are entirely separate; however, techniques like prompt engineering require the use of natural language, thereby significantly limiting the application of DNA large language models. This paper introduces a pre-trained model trained on the GPT-2 network, combining DNA sequences and English text, and uses a unified BPE tokenization method. We then convert classification and other downstream tasks into Alpaca format instruction data, and perform instruction fine-tuning on this pre-trained model to create a fine-tuned model capable of handling multiple tasks. The model has demonstrated its effectiveness in DNA related zero-shot prediction and multitask application. This research provides a highly promising direction for building a unified DNA sequence task framework.

## 1 Introduction

Large language models have emerged as a groundbreaking innovation in the field of artificial intelligence. Large language models are also applied in the analysis of DNA and protein sequences. Particularly for nucleic acid-focused tasks, such as DNABert2, HyenaDNA, ScBert *(1-3),* these studies primarily focus on issues such as the classification and structure prediction of DNA sequences. Similarly, for protein-related tasks, models like ProTrans, ProteinBERT, ESM2 have been developed, which include applications like structure prediction and function annotation *(4-6)* .

Most large DNA models use common large language model architectures and are pretrained using genomic DNA sequence data. For example, models like DNABERT, DNABERT2, and GEN-LM, which are based on the BERT-style architecture, use Masked Language Modeling (MLM) for training*(7,8)*. Models like DNAGPT, which are based on the GPT architecture, use Causal Language Modeling (CLM) for training*(9)*. Subsequently, based on the pretrained models, specific model heads are set up and fine-tuned for different downstream tasks. Large DNA language models perform better on tasks such as classification compared to smaller models like CNNs and LSTMs *(10-12)*.

Despite their successes, DNA large language models face several limitations. One significant challenge is the difficulty in applying novel prompt engineering techniques. Prompt engineering is the foundation of large model applications, as approaches like RAG , agents, and function calls all rely on well-crafted prompts to be built effectively.

For example, current large language models can easily accomplish the following tasks:

Question:

*Determine whether the sentiment of following text is positive or negative？*

*"""*

*This is the worst thing the TMNT franchise has ever spawned. I was a kid when this came out and I still thought it was deuce, even though I liked the original cartoon*

*"""*

Answer:

*negative*

However, for large language models specialized in DNA, the following tasks are extremely challenging.

Question:

*Determine the following dna sequence is promoter or terminator？*

*"""*

*GATTCCGTGGACTCGAGGCCCGCGTCCTCCGCCCTCCTGTGGCCCCGACCTGCCCGGAGCGCGTTCCCCGCCGGCGTCCGCTG CCGCTCACACCCACCCCAGTACCTGGCGGGCCCGGAGCGCGCGCG*

*"""*

Answer:

*Terminator*

In the context of large language model strategies, various downstream tasks such as translation, sentiment classification, named entity recognition, part-of-speech tagging, and more are unified under a single task: prompt-based dialogue.

But in the current exploration of large DNA models, a unifying "dialogue" task that can consolidate various downstream tasks has not yet been found. Instead, the approach still relies on fine-tuning the pre-trained model with small-scale, high-quality supervised data to create specialized models for specific tasks. In light of these limitations, there is a growing need to explore hybrid models that can integrate the strengths of DNA sequences with natural language.

For example, when large language models handle classification problems, a common approach is to introduce specific symbol labels, transforming the classification problem into a token prediction problem for the language model.

Consider two classification data points with categories 1 and 0:

*GATTCCGTGGACTCGAGGC (Category: 1)*

*CACCCACCCCAGTACCT (Category: 0)*

These can be converted into the following general text sequences:

*[CLS]GATTCCGTGGACTCGAGGC[SEP]1[SEP][CLS]CACCCACCCCAGTACCT[SEP]0*

Here:

*[CLS] (Classification Token):* this token is placed at the beginning of the input sequence, and it used as the representation of classification tasks.

[SEP] (Separator Token): Used to separate different sentences or paragraphs.

Such sequences can be treated like general text for training language models.

When predicting the category of following sequence:

*[CLS]GATTCCGTGGACTCGAGGC[SEP]*

The task is to predict the next token after [SEP].

Base on this idea, DNAGPT adds more types of tokens to the pre-training input sequences to represent concepts such as "categories" and "numerical values." For instance, <A> represents a positive example, and <N> represents a negative example. Additionally, it designs extra pre-training tasks to help the model learn to understand these concepts and prepare for downstream tasks *(9)*. HyenaDNA, on the other hand, introduces a "Soft Prompting" method that includes the SEP token to mark that the token to be predicted is a "category" rather than the "next base."*(2)*

In addition, gene annotation data has also been used for model training. For example, ESM3 introduced 256 functional keyword tokens *(13)*, while LucaOne utilized protein annotation information. In these models primarily focused on DNA analysis, the tokens mainly serve as classification IDs or label functions and generally do not utilize their semantic representation capabilities *(14)*.

On the other hand, many biological research papers contain sequences such as proteins. Studies like BioMedGPT and ProtSt leverage large amounts of biological literature and annotated protein sequences for unified large model training and fine-tuning *(15,16)*. This can effectively improve the performance of biomedical question answering, literature summarization, and natural language function prediction for protein sequences. However, these studies involve relatively little in terms of building a unified framework for biological sequence analysis tasks.

These innovative methods have achieved some SOTA results. But introducing a small number of specific natural language labels can also lead to issues with the label design system, label proliferation, and label semantic understanding.On the other hand, using a large amount of natural language text combined with a small amount of biological sequences makes it difficult to handle various downstream analysis tasks for biological sequences. There is still a gap before achieving a unified framework for downstream tasks.

However, these methods provide an interesting idea: since designing a complete system of task-specific symbols is challenging, why not use the full set of natural language symbols/tokens instead?

Our research involves using DNA sequences and English text to train a base GPT-2 model. We then convert the data related to DNA classification tasks into an instruction-tuning format and fine-tune the GPT-2 model accordingly. This process results in a model that can complete a variety of DNA-related tasks using natural language prompts.

## 2 Materials and methods

### 2.1 DNAHLM

To develop the DNAHLM, we adopted the GPT-2 as the foundation structure .We use English and DNA sequences as training corpora for pre-training, employing unified BPE encoding. Then, we convert DNA downstream tasks into standardized instruction data to fine-tune the pre-trained model. The training process is shown in Fig.1:

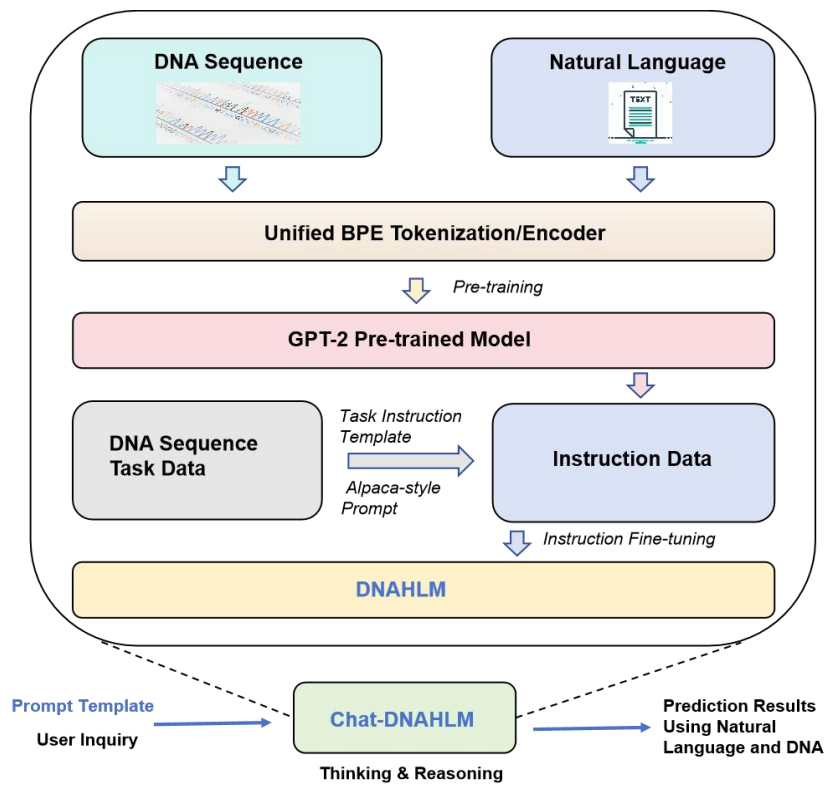

Fig.1　DNAHLM training process. This model uses human genome data and English encyclopedia data as pre-training corpora, employing the BPE method for uniform encoding without distinguishing between the two types of text. It then undergoes training from scratch based on the GPT2-small network, resulting in a pre-trained model. Following this, specific prompt templates are used to convert DNA downstream tasks into instructions. These instructions are then formatted into the standard Alpaca format, generating instruction data for fine-tuning the pre-trained model. This process leads to the creation of DNAHLM. The usage of this model can adopt a dialogue mode based on prompt engineering.

The technology for training GPT-2 base models from scratch is already very mature. We will primarily focus on the fine-tuning methods. For comparison, we used two fine-tuning methods: classification tuning and instruction tuning.

1 Classification tuning.This is currently the most common application approach for large models in the field of DNA. In classification tuning, the model is trained to recognize specific category label ids, such as "0(promoter)" and "1(terminator)." Classification tasks can also include recognizing Splice site, Transcription factor, etc. However, a model fine-tuned for classification can only make judgments about the specified categories and cannot perform other types of judgments on the input text, Fig.2. For example, promoter detection need one classification fine-tuned model, the Transcription factor classification need another.

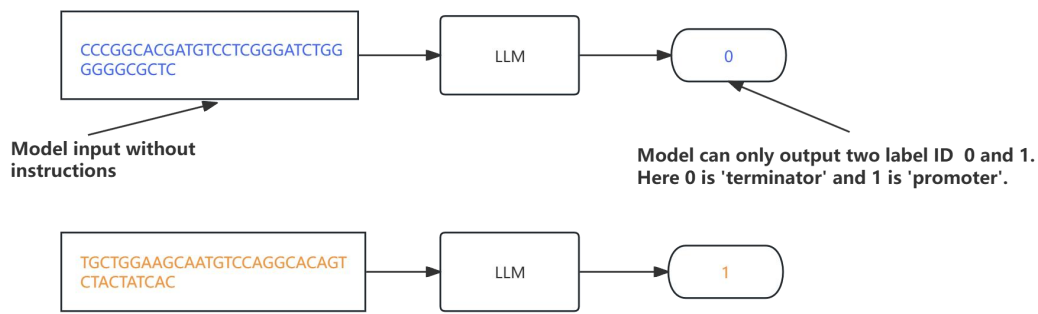

Fig2 DNA Classification fine-tuning tasks.The input to the model is a DNA sequence, and the output is a list of IDs.

2 Instruction tuning. Instruction tuning involves training the model on specific tasks to enhance its ability to understand and execute tasks described in natural language prompts, as shown in Fig.3. We can think of classification-tuned models as highly specialized models. By contrary，Models fine-tuned with instruction tuning are generally capable of performing a broader range of tasks. That means we could predict the DNA Splice site, Transcription factor, even functions by one model. This is precisely the focus of our research.

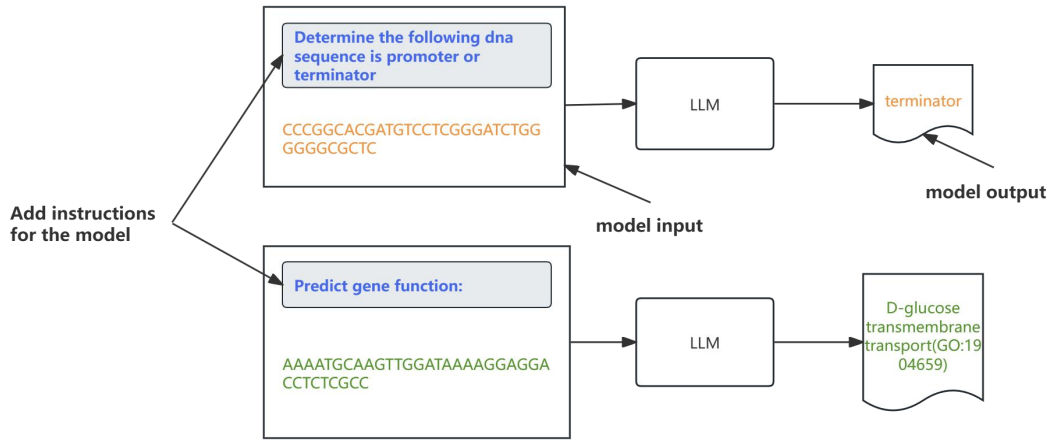

Fig 3. DNA Instruction fine-tuning tasks. The input to the model is instruction data, typically composed of DNA sequences and natural language text. The output is also in this format. Due to the adoption of a unified encoding method, the model can support the promoter classification task mentioned above, as well as the gene function prediction task below.

The DNAHLM uses the exact same model architecture as GPT-2, with the primary difference being the training corpus. For the research validation, the base model is structured as GPT-2 Small, which can be trained on a single 4090 GPU. For classification fine-tuning, a typical classification head is used, outputting two classes. For instruction fine-tuning, the same head as the petrained model is used, and the training mode is also identical to that of the pre-training. The structure of DNAHLM is shown in Fig.4.

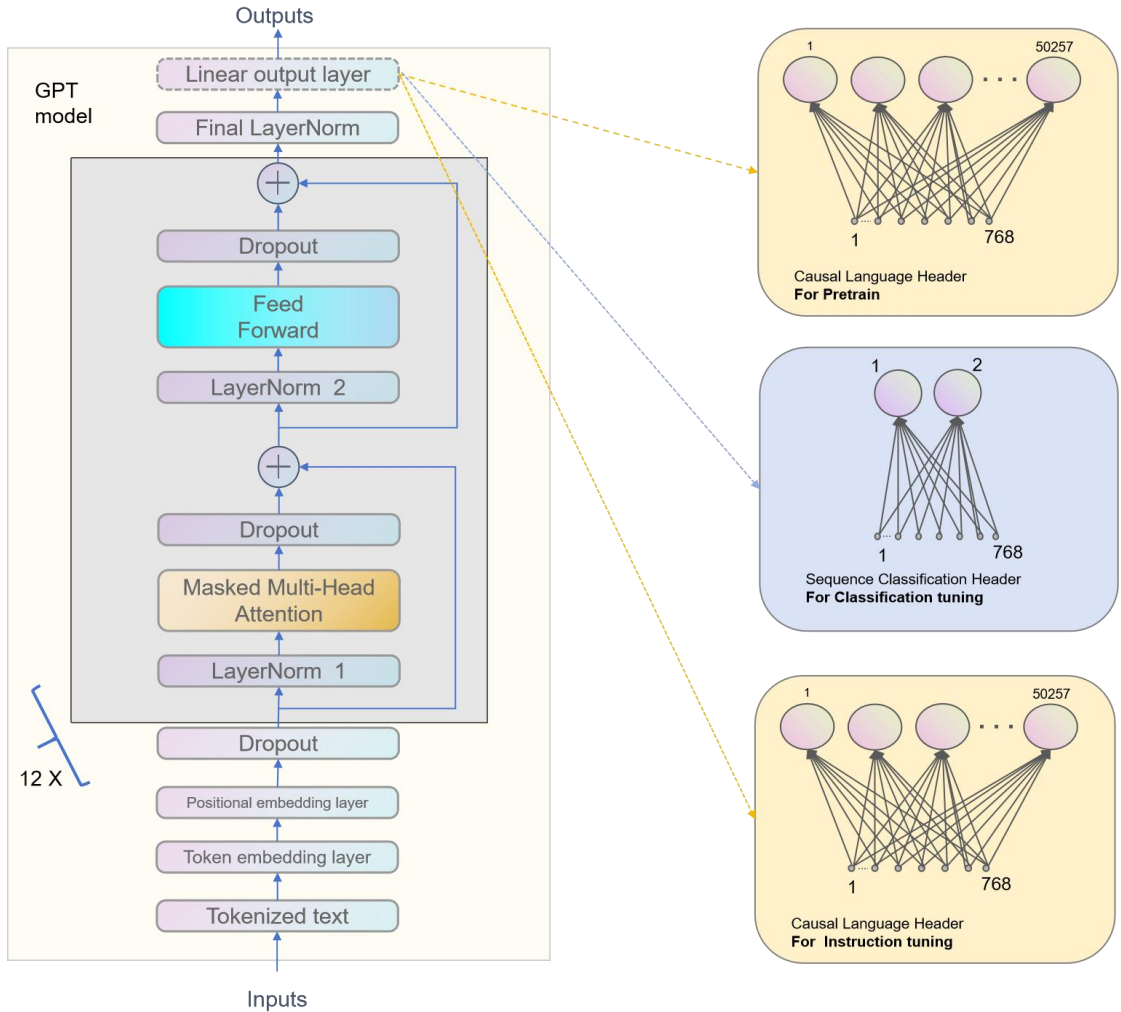

Fig.4. Structure of DNAHLM.The model architecture is identical to that of GPT-2 small. Pre-training and instruction tuning use the same causal language header,the layer mapped 768 hidden units to 50,257 units (the number of tokens in the vocabulary). Classification uses sequence classification header, The layer maps from 768 hidden units to only 2 units, where the 2 units represent the two classes id.

## 2.2 Data Preprocessing

To prepare the data for training the DNAHL model, we first need to convert DNA and English sequences into a format suitable for processing. This involves several steps:

1 DNA and English datasets. For the DNA data, we used human genome data, dividing the human genome into segments of 300 to 1000 base pairs (bp) in length, and then randomly selected some of these segments as the DNA training data. For the English text data, we used data from the English version of Wikipedia. We selected 150MB of DNA sequence data and 150MB of English text data, and then combined them to serve as the training data for the GPT-2 model.

2 Fine-tuning datasets. For fine-tuning, we used typical DNA sequence analysis tasks and converted them into a fine-tuning data format for instruction tuning. A typical example includes the following downstream tasks:

- Core Promoter Detection

- Transcription Factor Prediction

- Promoter Detection

- Splice Site Detection

Since there is currently a lack of a standard evaluation dataset for DNA downstream tasks, we used the GUE datasets constructed in the DNA-BERT paper and the GENA-LM Promoter prediction datasets *(1,8)*. The original data was hosted on Google Drive; we performed a simple format conversion and uploaded it to Hugging Face *(17)*.

These tasks were transformed into an instruction-tuning format to fine-tune the model.

Instruction tuning is a supervised learning method where the training data consists of instructions, inputs, and outputs. These three components are then formatted into a specific prompt format. Below are two common methods of formatting.Fig.5. In this article, we will use the Alpaca-style prompt formatting method.

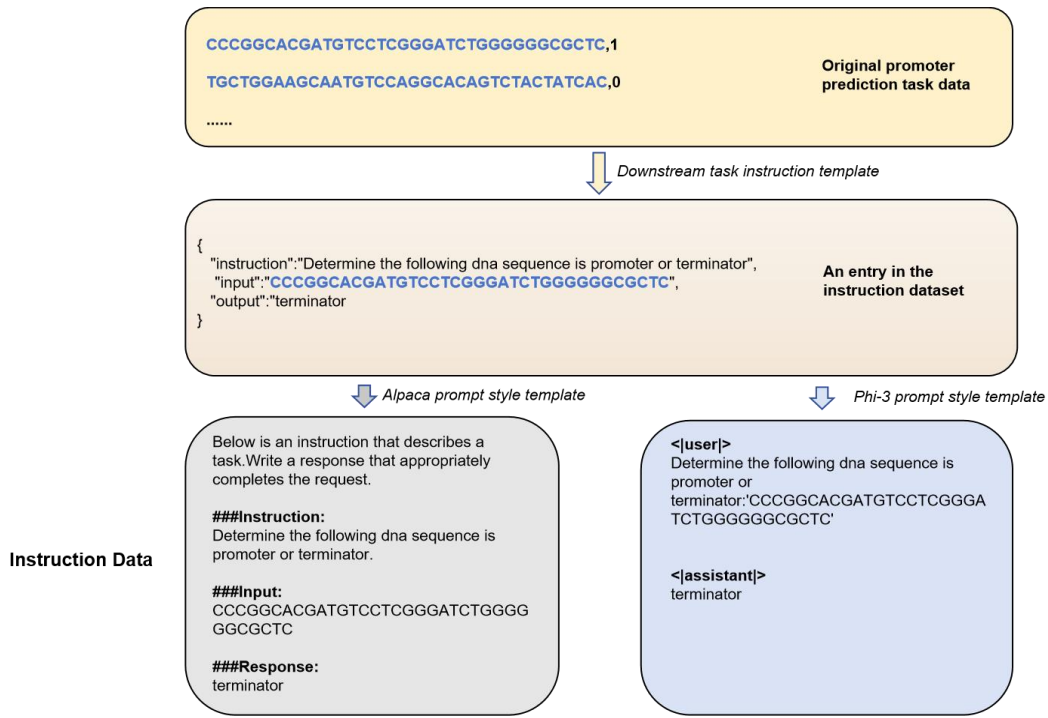

Fig.5 Build instruction datasets. For different downstream tasks, design specific instruction templates to convert downstream task data into instructions. Then, using the Alpaca prompt template, transform these instructions into uniformly formatted instruction fine-tuning data.

Currently, the main downstream tasks for DNA are classification tasks. A typical classification task is shown in the table below:

Table.1 An example of a classification task

| Task name | sequence | label name | label |
| --- | --- | --- | --- |
| promoter detection | CATGCGGGTCGATAT... | Non-promoter | 0 |
| promoter detection | CTGAATCTCTCAGCC... | promoter | 1 |

We use tabular data as an example to illustrate the entire process of converting DNA downstream tasks into instruction fine-tuning data:

**Step 1.** For classification tasks, a template like the following can generally be used to construct instructions：

*instruction = "Determine the {task_name} of following dna sequence, The result will be one of the following: {label_name1}, {label_name2}, {...}"*

**Step 2.** Correspondingly, a classification data can be converted into an instruction tuning data：

*{'instruction': 'Determine core promoter detection of following dna sequence, The result will be one of the following: Non-promoter, promoter.',*

*'input': 'CATGCGGGTCG...',*

*'output': 'Non-promoter'}*

**Step 3**. By applying the Alpaca prompt template, the text data for fine-tuning training can be formed：

*Below is an instruction that describes a task. Write a response that appropriately completes the request.*

*### Instruction:*

*Determine core promoter detection of following dna sequence, The result will be one of the following: Non-promoter, promoter.'*

*### Input:*

*TCTTTCTCTTCTGTATCATTCTACTT...*

*### Response:*

*Non-promoter*

It's important to note that the instruction conversion templates for downstream tasks are not fixed. For example, a binary classification problem can also use the following template:

*Determine whether the {task_name} of following DNA sequence {sequence} answer with {label_name1} or {label_name2}*

Then the instruction tuning data will be:

*{'instruction': 'Determine the core promoter detection of following DNA sequence 'CATGCGGGTCG...', answer with Non-promoter or promoter .',*

*'input': '',*

*'output': 'Non-promoter'}*

Here, you only need to set the input as empty. In the instructions, DNA sequences and English text can be combined arbitrarily, which allows our model to be compatible with a wider range of downstream tasks.

## 2.3 Training Strategy

### 2.3.1 Tokenization and Encoding

Currently, most large biological models that support multimodal data generally use different encoding methods for annotations, DNA, proteins, structural information, and natural language. For instance, BioMedGPT uses GraphMVP as the encoder for 2D molecular graphs, ESM2-3B as the encoder for protein sequences, and BPE tokenization for encoding natural language (*15*). LucaOne model distinguishes nucleotides and amino acids by utilizing token-type encoding, assigning 0 to nucleotides and 1 to amino acids (*14*).

These approaches, where different sequences use different encoders, can more effectively capture the feature information of the sequences. However, it introduces an encoder selection issue and requires preliminary determination of the sequence types when handling downstream tasks. This method increases the difficulty of building a unified model for DNA tasks.

The GPT-2 model uses BPE (Byte Pair Encoding) by default to encode natural language.

For DNA sequences, K-mer tokenization an approach that has been widely used. The k-mer representation incorporates richer contextual information for each deoxynucleotide base by concatenating it with its following ones. The concatenation of them is called a k-mer. For example, a DNA sequence 'ATGGCT' can be tokenized to a sequence of four 3-mers: {ATG, TGG, GGC, GCT} or to a sequence of two 5-mers: {ATGGC, TGGCT}. But Byte-Pair Encoding (BPE) tokenization has been shown to be more efficient in large DNA models. For example, the sequence 'ATGGCT' can be tokenized as {ATGG, CT}.

Therefore, we also used the BPE tokenization algorithm. We built a vocabulary of approximately 50,000 tokens based on a mixed corpus of English and DNA sequences, with DNA tokens comprising about 55% and English tokens about 45%. The tokenizer includes special tokens::<|endoftext|>. The statistical distribution of the dictionary is shown in Fig.5.

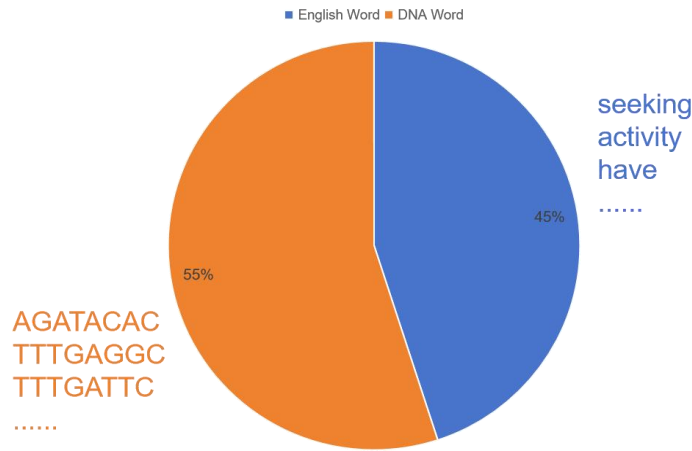

Fig.6 Statistical distribution of the dictionary. Among these, English vocabulary accounts for 45%, and DNA vocabulary accounts for 55%.

**2.3.2 Pre-training**

According to the typical design of GPT-2, it accepts sequences with a maximum length of 1024 as input. We used the same architecture as the GPT-2 Small model, which consists of 12 Transformer layers, each with 768 hidden units and 12 attention heads. This small model has approximately 117 million parameters. We trained the GPT-2 model using mixed-precision floating-point arithmetic on a machine equipped with a single Nvidia 4090 GPU. We employed a dynamic learning rate schedule, and the model was trained for a total of 3 to 5 epochs.

**2.3.3 Classification Fine-tuning**

For each downstream application, we started from the pre-trained parameters and fine-tuned DNAHLM with task-specific data. The model's header is set to the commonly used softmax classification form; see Fig. 4 for specifics. We utilized the same training tricks across all the applications, where the learning rate was first linear warmed-up to the peak value and then linear decayed to near 0. Due to the limited amount of data, we divided the fine-tuning datasets into a training set and a validation set and did not set aside a dedicated test set. Model evaluations were conducted on the validation set.

Due to the smaller scale of model parameters and the datasets size being in the tens of thousands of samples, we used the full-parameter fine-tuning method. Fine-tuning can be completed within about an hour on a single 4090 GPU. If using larger models such as GPT-2 Large or datasets with greater volumes, the LoRA method can be adopted, combined with the DeepSpeed multi-GPU mode for fine-tuning.

### 2.3.4 Instruction Fine-tuning

The method for instruction tuning is entirely consistent with the training method of the GPT-2 pre-trained model. All use exactly the same causal language model head; see Fig. 4 for specifics. This involves treating the constructed instruction data as general text sequences and inputting them into the GPT-2 model. In other words, instruction tuning is the same as the pre-training process, which is key to enabling multi-task handling. This is also key to how ChatGPT-like models are able to handle a variety of natural language tasks. We typically train for 2 to 3 epochs. Since the training data is relatively small, fine-tuning on a 4090 GPU usually takes only about ten minutes to complete.

### 2.4 Model Evaluation

The performance of the DNAHL model is evaluated using a series of benchmarks that assess its capabilities in both DNA sequence and promotion tasks.

For classification fine-tuning, we evaluate the model according to typical classification tasks. This benchmark evaluates the model's ability to classify DNA sequences into different categories based on their functional or structural properties. The model's accuracy is used to measure its performance.

For the evaluation of the instruction-tuned model, as a comparison, we commonly use accuracy as evaluation metrics. The assessment is based on whether the model's output tokens semantically match the expected ones. In contrast, the output for classification fine-tuning is typically a numerical class ID, such as 1, 2, etc, we need to output exact equality.

For example, if the model outputs "promoter AGCC GGG" while the expected output is "promoter ", we still consider the model's prediction to be accurate. This is because, as a generative model, GPT is not specifically trained to recognize stop tokens. With more training data and a larger model size, it would be possible to produce outputs that exactly match the expected tokens.

## 3 Experimental Results

For the specific promoter prediction task, the classification fine-tuned model achieved an accuracy of approximately 76%-83%, while the instruction-tuned model achieved an accuracy of approximately 74%. This indicates that the instruction-tuned model remains effective even when dealing with specialized classification problems.

For multiple tasks such as Core Promoter Detection and Transcription Factor Prediction, the classification fine-tuning method requires training a different model for each task, with an average accuracy of approximately 80%. In contrast, the instruction-tuned model, when facing multiple types of classification tasks, achieved an accuracy of around 83%, but these results were obtained using a single model.

Table.2. Accuracy of different models

| Task | DNAHLM(Classification Fine-tuning) | GENA-LM | DNABert2 | DNAHLM(Instruction Fine-tuning) |
|---|---|---|---|---|
| Promoter Detection 1 | 0.83 | 0.76 | - | 0.74 |
| Core Promoter Detection | 0.76 | - | 0.74 | 0.83 |
| Transcription Factor Prediction | 0.75 | - | 0.87 | |
| Promoter Detection 2 | 0.88 | - | 0.94 | |
| Splice Site Detection | 0.82 | - | 0.86 | |

Here, 'Promoter Detection 1' uses data from the GENA-LM paper, while the rest use data from the DNABert2 paper.

By contrast, the current state-of-the-art (SOTA) model DNABert2 has an average accuracy of approximately 0.85, which is slightly higher than our instruction fine-tuned model. However, this also validates the effectiveness of our method, which involves unified encoding of different sequence types and fine-tuning the model using a standardized instruction format.

More importantly, the instruction-tuned model can interact with users in a conversational manner and can leverage prompt engineering, RAG, chain of thought, and other techniques commonly used with large language models.Fig7. This significantly expands the range of applications for the model.

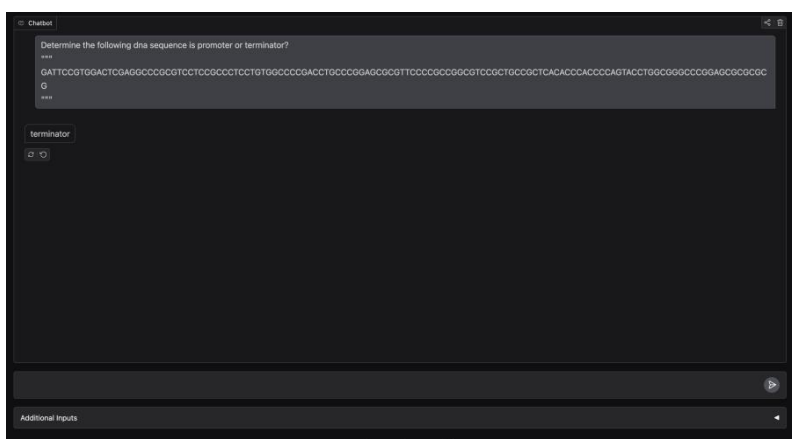

Fig.7. Chat with DNAHLM

## 4 Conclusion

In summary, this research paper presents the development and evaluation of the DNAHL model, a hybrid large language model that integrates DNA sequences with human language.

The main innovations of this study are twofold. First, it uses unified BPE encoding for both DNA sequences and English text. Then, using prompt templates, it converts different DNA downstream tasks into standardized instruction fine-tuning data. Although this method ignores the biological characteristics of DNA, it still yields good results in downstream tasks such as Promoter prediction.

This "simple" approach to model construction has clear advantages. Similar to ChatGPT, it enables solving various DNA downstream tasks through a conversational interface, where the input can be any combination of DNA sequences and natural language. Additionally, it allows for the direct application of mature large-model application frameworks such as prompt engineering, RAG (Retrieval-Augmented Generation), function calls, and agents to develop large-scale DNA model applications. The complete training code for the model is open-sourced on GitHub *(18)*.

Future research could involve using larger-scale models, such as llama series, and training on more extensive datasets, such as genomic and protein data from multiple model organisms , additional fine-tuning data for various DNA-related downstream tasks.